\documentclass{PoS}
\usepackage{mathtools}
\usepackage{xargs}                                   % Handy way to define commands with several default arguments
\usepackage{xparse}                                  % Handy way to define star version of commands
\usepackage{ifthen}                                  % provides \ifthenelse test  
\usepackage{xifthen}                                 % provides \isempty test
\usepackage[binary-units, per-mode = symbol, exponent-product = \cdot]{siunitx}
\usepackage{array,booktabs}
\usepackage[]{cleveref}                              % It has to be loaded AFTER hyperref!

\crefname{figure}{Figure}{Figures}
\crefname{table}{Table}{Tables}

\newcommand{\order}[2][]{\mathcal{O}\mathopen#1(#2\mathclose#1)}

\makeatletter
  \newcommandx{\intd}[4][1=, 2=, 3=]{
    \ifthenelse{\isempty{#1} \AND \isempty{#2}}{
        \int_{#1}^{#2}\mathrm{d}^{#3}#4\mspace{4mu}\@ifnextchar\d{\mspace{-4mu}}{}
    }{
        \int_{#1}^{#2}\mspace{-6mu}\mathrm{d}^{#3}#4\mspace{4mu}\@ifnextchar\d{\mspace{-4mu}}{}
    }
  }
  \newcommand{\intD}[1]{\int D#1\mspace{4mu}\@ifnextchar\D{\mspace{-4mu}}{}}
  \renewcommand{\d}[2][]{\mspace{4mu}\mathrm{d}^{#1}#2\mspace{4mu}\@ifnextchar\d{\mspace{-4mu}}{}}
  \newcommand{\D}[1]{\mspace{4mu}D#1\mspace{4mu}\@ifnextchar\D{\mspace{-4mu}}{}}
\makeatother

\newcommandx{\group}[3][2=]{
    \ifthenelse{\isempty{#2}}{
        \ifthenelse{\isempty{#3}}{\MakeUppercase{#1}}{\MakeUppercase{#1}(#3)}
    }{
        \ifthenelse{\isempty{#3}}{\MakeUppercase{#1}_{#2}}{\MakeUppercase{#1}_{#2}(#3)}
    }
}
\DeclarePairedDelimiter\abs{\lvert}{\rvert}%
\DeclarePairedDelimiter\norm{\lVert}{\rVert}%
\makeatletter
\let\oldabs\abs
\def\abs{\@ifstar{\oldabs}{\oldabs*}}
\let\oldnorm\norm
\def\norm{\@ifstar{\oldnorm}{\oldnorm*}}
\makeatother
\DeclareFontFamily{U}{matha}{\hyphenchar\font45}
\DeclareFontShape{U}{matha}{m}{n}{
      <5> <6> <7> <8> <9> <10> gen * matha
      <10.95> matha10 <12> <14.4> <17.28> <20.74> <24.88> matha12
      }{}
\DeclareSymbolFont{matha}{U}{matha}{m}{n}
\DeclareMathSymbol{\to}{3}{matha}{"D1}

\newcommandx{\action}[2][1=, 2=]{
    \ifthenelse{\isempty{#2}}{
        \ifthenelse{\isempty{#1}}{S}{S_{#1}^{\vphantom{X}}}
    }{
        \ifthenelse{\isempty{#1}}{S_{\vphantom{X}}^{#2}}{S_{#1}^{#2}}
    }
}

\newcommand{\mass}[1][0]{m_{#1}}
\newcommandx{\field}[7][1=\:, 3=, 5= , 7=\:]{#1#2^{#3}_{#4}\mathopen#5(#6\mathclose#5)#7}  % Generic field
\newcommand{\PartitionFunction}{Z}
\newcommandx{\PF}[4][1=,2=,4=\big]{
    \ifthenelse{\isempty{#3}}{
        \PartitionFunction^{#1}_{#2}
    }{
        \PartitionFunction^{#1}_{#2}\mathopen#4(#3\mathclose#4)
    }
}
\newcommandx{\ev}[3][2=, 3=]{\mathopen#3\langle#1\mathclose#3\rangle_{#2}}

\def\amd{\texttt{AMD}}
\newcommand{\clqcd}{\texttt{CL\kern-.15em\textsuperscript{2}QCD}}
\newcommand{\clqcdUrl}{\href{url}{\texttt{https://github.com/CL2QCD/cl2qcd}}}
\def\CP{Columbia plot}

\def\gpu{\texttt{GPU}}

\def\icp{imaginary chemical potential}
\def\lcsc{\mbox{\texttt{L-CSC}}}

\def\ocl{\texttt{OpenCL}}
\def\PL{Polyakov loop}
\def\pd{phase diagram}
\def\picp{purely \icp}
\def\qcd{QCD}

\def\RW{\mbox{Roberge-Weiss}}
\newcommand{\nD}[1]{#1D}
\def\threeD{\nD{3}}

\NewDocumentCommand{\First}{s}{%
    \IfBooleanTF{#1}{1$^{\text{st}}$}{first}%
}
\NewDocumentCommand{\Second}{s}{%
    \IfBooleanTF{#1}{2$^{\text{nd}}$}{second}%
}
\NewDocumentCommand{\Iorder}{s}{%
    \IfBooleanTF{#1}{\First-order}{\First{} order}%
}
\NewDocumentCommand{\IIorder}{s}{%
    \IfBooleanTF{#1}{\Second-order}{\Second{} order}%
}

\def\Ns{N_{\sigma}}
\def\Nt{N_{\tau}}
\def\Nf{N_f}
\newcommand{\StaggPh}[2][]{\ifthenelse{\isempty{#1}}{\eta_{#2}}{\eta_{#2}(#1)}}

\newcommand{\latSpacing}{a}

\newcommandx{\Poly}[2][1=, 2=]{\ifthenelse{\isempty{#1}}{L^{#2}}{L^{#2}(#1)}}

\newcommand{\PolyIm}{L_{\text{Im}}}
\newcommand{\muRe}[1][]{\ifthenelse{\isempty{#1}}{\mu}{\mu_{#1}}}
\newcommand{\muIm}{\mu_i^{\vphantom{x}}}
\newcommand{\muImRW}{\mu_i^{\text{\textsc{rw}}}}

\newcommand{\mud}{m_{u,d}}

\newcommand{\mpi}{m_{\pi}}
\newcommand{\mTric}[1][]{m_{#1}^{\text{tric}}}

\newcommand{\betaC}{\beta_{\text{c}}}
\newcommand{\kurtosis}{B_4}

\graphicspath{{Figures/}}

\title{Roberge-Weiss transition in $\mathbf{\Nf=2}$ \qcd{} with staggered fermions and $\mathbf{\Nt=6}$}

\ShortTitle{Roberge-Weiss transition with two staggered flavors and $\Nt=6$}

\author{Owe Philipsen\\ 
        Institut f\"ur Theoretische Physik - Johann Wolfgang Goethe-Universit\"at\\
        Max-von-Laue-Str. 1, 60438 Frankfurt am Main\\   
        E-mail: \email{philipsen@th.physik.uni-frankfurt.de}
}

\author{\speaker{Alessandro Sciarra}\\ 
        Institut f\"ur Theoretische Physik - Johann Wolfgang Goethe-Universit\"at\\
        Max-von-Laue-Str. 1, 60438 Frankfurt am Main\\
        E-mail: \email{sciarra@th.physik.uni-frankfurt.de}
}

\abstract{The \qcd{} phase diagram at imaginary chemical potential exhibits a rich structure and studying it can constrain the phase diagram at real values of the chemical potential.
          Moreover, at imaginary chemical potential standard numerical techniques based on importance sampling can be applied, since no sign problem is present.
          In the last decade, a first understanding of the \qcd{} \pd{} at \picp{} has been developed, but most of it is so far based on investigations on coarse lattices ($\Nt=4$, $\latSpacing=\SI{0.3}{\femto\meter}$).
          Considering the $\Nf=2$ case, at the \RW{} critical value of the \icp, the chiral/deconfinement transition is \Iorder{} for light/heavy quark masses and \IIorder{} for intermediate values of the mass: there are then two tricritical masses, whose position strongly depends on the lattice spacing and on the discretization.
          On $\Nt=4$, we have the chiral $\mpi^{\text{tric.}}=\SI{400}{\mega\electronvolt}$ with unimproved staggered fermions and $\mpi^{\text{tric.}}\gtrsim\SI{900}{\mega\electronvolt}$ with unimproved pure Wilson fermions.
          Employing finite size scaling we investigate the change of this tricritical point between $\Nt=4$ and $\Nt=6$ as well as between Wilson and staggered discretizations.
}

\FullConference{The 34th annual International Symposium on Lattice Field Theory\\
                24-30 July 2016\\
                University of Southampton, UK
}

\begin{document}

\section{Introduction}

The structure of the phase diagram of \qcd{} at zero chemical potential has been studied for more than a decade by now.
Investigations both in the chiral ($\mass[]\to0$) and quenched ($\mass[]\to\infty$) limits are interesting on their own, since they allow to understand the influence of confinement and chiral symmetry breaking on the thermal transition.
The structure of the \CP{} which emerges from studies on very coarse lattices~\cite{deForcrand:2007rq,Bonati:2014kpa,Saito:2011fs} is qualitatively the same for different formulations and lattice spacings: The order of the phase transition for light and heavy quarks is \Iorder, while it becomes a crossover for intermediate values of the mass.
Nevertheless, quite significant discrepancies regarding the position of the $\group{z}[2]{}$ lines separating these regions have been found using different fermions discretizations and no continuum extrapolation of any feature is available.

\begin{figure}[b]
    \centering
    \includegraphics[width=0.7\textwidth]{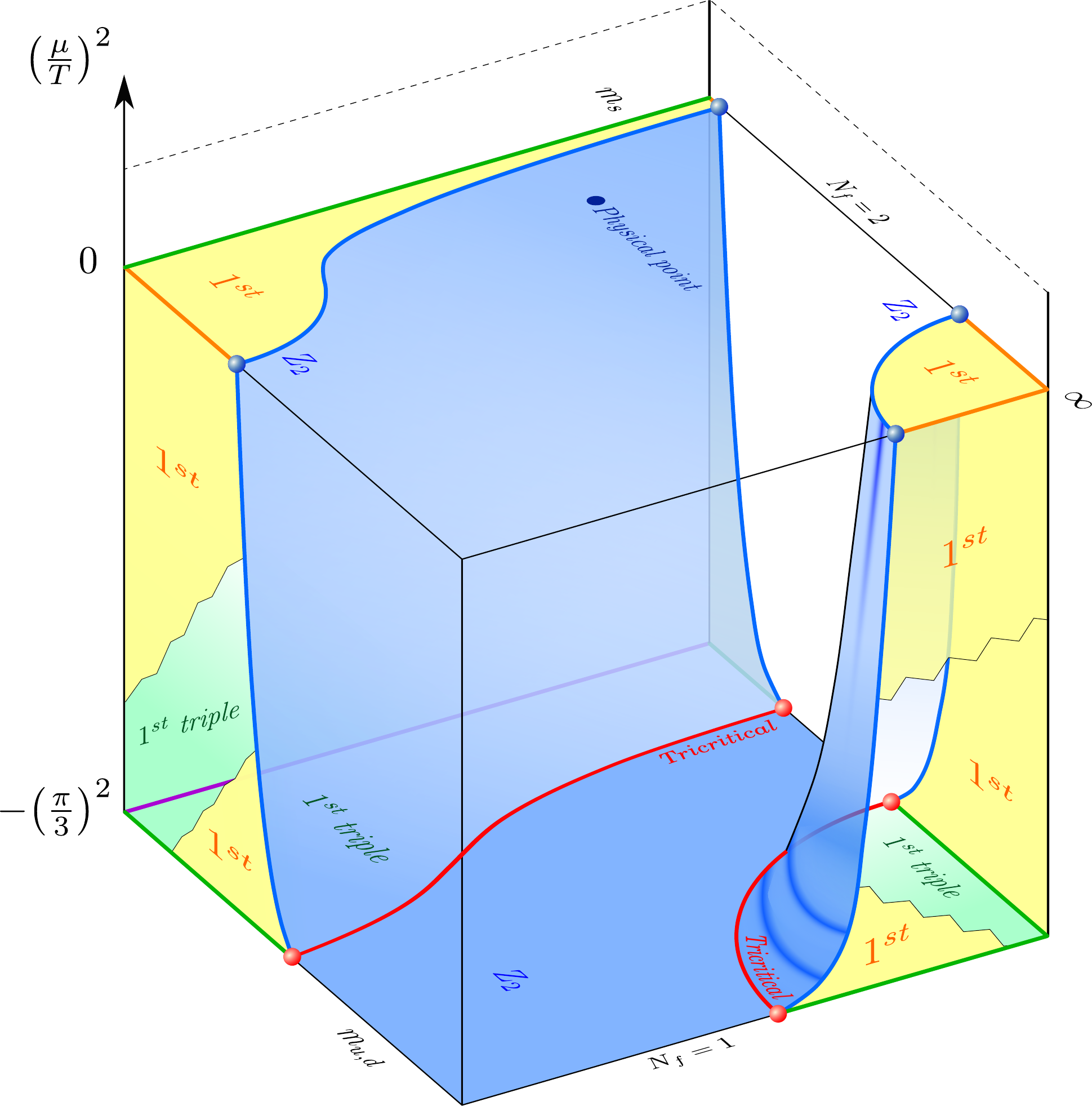}
    \caption{Possible scenario of the \threeD{} \CP{}.
             The region below the $\muRe=0$ plane is free from the sign problem and it can be directly
             studied.
             The \Iorder*{} regions on the front vertical planes have been interrupted, in order to allow
             a better understanding of the picture.}
    \label{fig:3Dplot}
\end{figure}

Introducing a \picp{} and varying it between $\muIm=0$ and the \RW{} critical value, the \Iorder*{} regions (crossover region) of the \CP{} continuously enlarge (shrinks), ending in \Iorder*{} triple (\IIorder*) regions at  $\muIm=\muImRW=T\cdot\pi/3$, because of the \RW{} symmetry,
\begin{equation*}
     \PF{\frac{\muIm}{T}}[\Big]=\PF[\prime]{\frac{\muIm}{T}+\frac{2\pi k}{3}}[\Big]\;.
\end{equation*}
The $\group{z}[2]{}$ lines change their nature as well, becoming tricritical lines. 
In \cref{fig:3Dplot}, a possible scenario of the so-called \threeD{} \CP{} has been drawn.
Even though it corresponds to the findings of previous studies, on finer lattices, the chiral \Iorder*{} region seems to shrink and it has not been ruled out that it could disappear in the continuum limit, with a consequent \IIorder*{} phase transition for massless quarks~\cite{deForcrand:2016pos}.

Focusing on the \RW{} plane with $\Nf=2$ degenerate quarks, the position of the tricritical points separating the \Iorder*{} triple regions from the $\group{z}[2]{}$ one has been found in the last years on $\Nt=4$ lattices both with unimproved Wilson fermions~\cite{Philipsen:2014rpa} and with unimproved staggered fermions~\cite{Bonati:2010gi}.
In the first case, this study has been already repeated on $\Nt=6$, finer lattices~\cite{Cuteri:2015qkq} and a comparison between the values of the light tricritical pion masses on the coarser and finer lattices shows that cut-off effects are still large.
It is interesting to observe that the chiral $\mpi^{\text{tric.}}=\SI{400}{\mega\electronvolt}$ with unimproved staggered fermions found on $\Nt=4$ lattices is already lighter than that found on $\Nt=6$ lattices ($\mpi^{\text{tric.}}=\SI{}{\mega\electronvolt}$) with unimproved staggered fermions.
A better understanding is clearly needed and having the values of $\mpi^{\text{tric.}}$ on a finer lattice with staggered fermions is a first step in this direction.

\section{Simulation details}

A standard way to simulate $\Nf=2$ degenerate flavors of staggered fermions is to use the RHMC algorithm~\cite{Kennedy:1998cu}.
In the present study we used a \gpu{} implementation of it present in the \clqcd{} software, a \ocl{} based code which is publicly available\footnote{~\clqcdUrl} and that has been optimised to run efficiently on \amd{} graphic cards~\cite{Philipsen:2014mra}.
All our simulations were run on the \lcsc{} cluster~\cite{RohrBNLPP15} in Darmstadt.

Our numeric setup, aside from the different fermion discretization, is completely analogous to that used in~\cite{Cuteri:2015qkq}.
The fourth standardized central moment of the imaginary part of the (spatially averaged) \PL,
\begin{equation*}
    \kurtosis(\PolyIm)\equiv\frac{\ev{(\delta \PolyIm)^4}[][\big]^{\phantom{2}}}
                     {\ev{(\delta \PolyIm)^2}[][\big]^{2}}
    \qquad\text{with}\qquad \delta \PolyIm \equiv \PolyIm - \ev{\PolyIm} \;,
\end{equation*}
also known as kurtosis, has been used to identify the nature of the \RW{} end- or meeting point.
Working at $\muImRW/T=\pi$ (i.e. on the second \RW{} plane), $\PolyIm$ can be used as order parameter
and its kurtosis is expected to vary from $3$ (crossover) for low temperature to $1$ (\Iorder) for high temperature.
On finite spatial volumes, $\kurtosis$ will be a smooth function, becoming discontinuous only in the thermodynamic limit,
\begin{equation*}
    \kurtosis(\beta)=2\,\Theta(\betaC-\beta)\;.
\end{equation*}   
Close enough to the critical temperature, if finite size effects are negligible, the kurtosis of $\PolyIm$ is a function of $x\equiv(\beta-\betaC)\,\Ns^{1/\nu}$ only, where $\nu$ is a critical exponent associated to the phase transition.
Therefore,
\begin{equation*}
    \kurtosis(\beta, \Ns)=\kurtosis(\betaC,\infty) + a_1\,x + \order{x^2}
\end{equation*}
and the order of the phase transition can be understood performing a linear fit of our data and looking at $\nu$, which takes its universal value depending on the transition. 
 
\begin{figure}[t]
    \centering
    \includegraphics[width=0.45\textwidth]{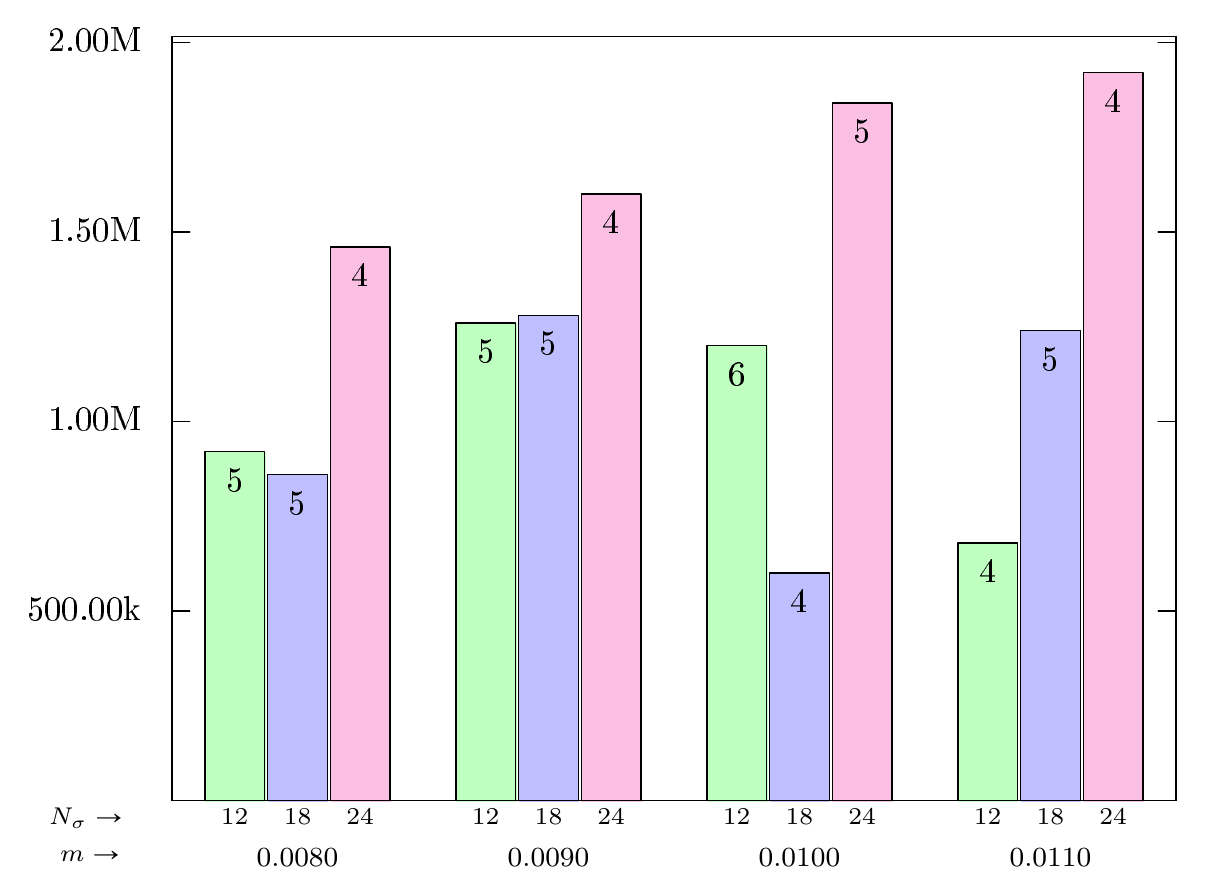} \qquad
    \includegraphics[width=0.45\textwidth]{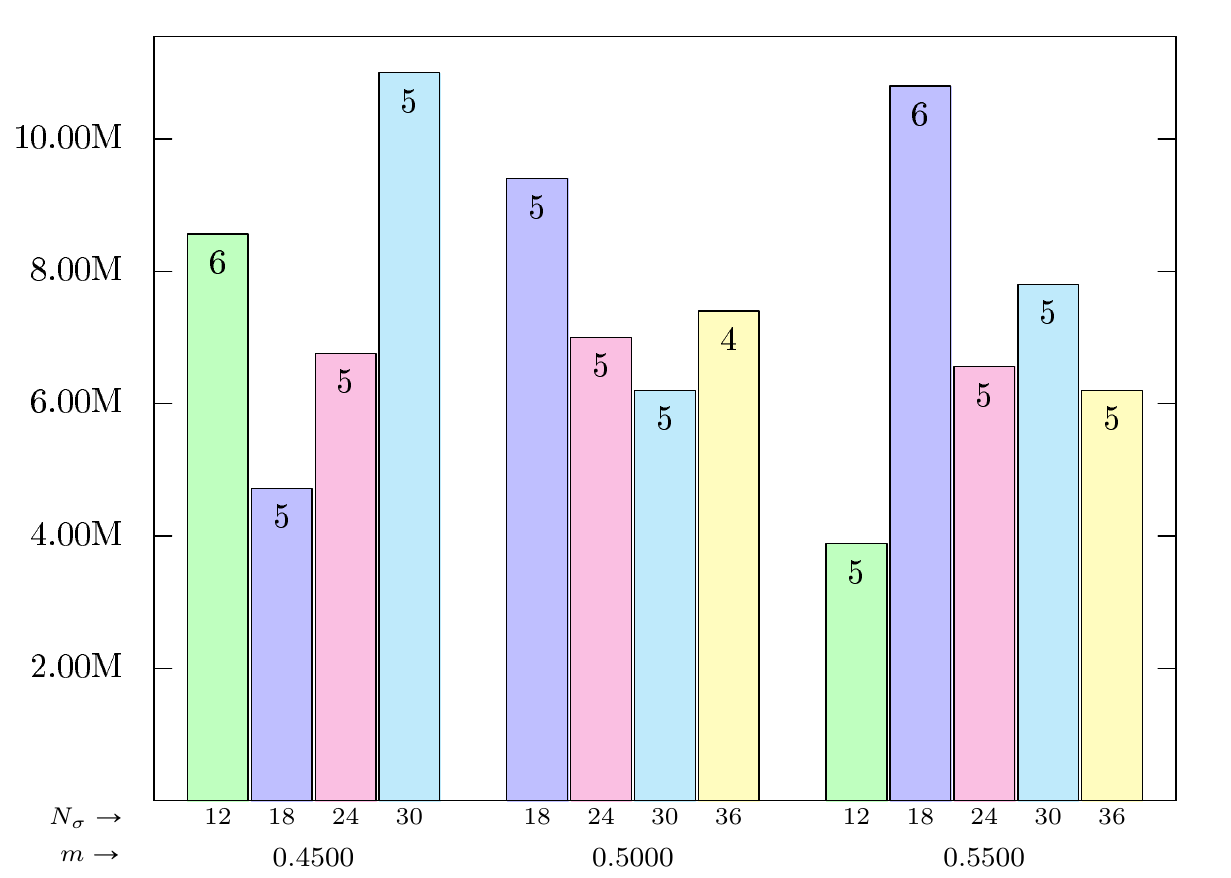}
    \caption{Example of collected statistics per volume at some of the simulated masses.
             The number written inside each histogram bar indicates the number of simulated $\beta$ around the critical one for the given volume.
             Observe the different vertical scales in the light and heavy regions.}
    \label{fig:statistics}
\end{figure}

In order to locate the two tricritical points, a scan in mass is needed.
For each value of $\mud$, we simulated at a fixed temporal lattice extent $\Nt=6$ and at a fixed value of the chemical potential $\latSpacing\muImRW=\pi/6$.
In the finite size scaling analysis, the three larger simulated volumes have been used, always with $\Ns\ge12$ (up to $\Ns=36$).
For each lattice size, $3$ to $8$ values of $\beta$ around the critical temperature have been simulated, each with $4$ Markov chains.
The accumulated statistics per $\beta$ has not been constant since we adopted two different precisions to decide when to stop our simulations in the light and in the heavy regions.
In particular, for large (small) masses we required the kurtosis of the imaginary part of \PL{} to be the same on all the chains within $2$ ($3$) standard deviations.
Since this condition can be met for poor statistics because of large errors, as rough rule of thumb, we required all the $4$ values of the kurtosis at a given temperature to span with their error an interval not wider than 0.5.
In \cref{fig:statistics}, the total number of unitary trajectories produced per volume at $8$ of the $18$ simulated mass values is reported.

Since to understand the order of the phase transition at a given value of the mass is not immediate (both in terms of time and numerical resources), previous studies~\cite{deForcrand:2007rq,Bonati:2010gi} can help in choosing how to perform the scan in $\mud$.
In particular, at $\muRe=0$ with $3$ degenerate flavors of staggered fermions, it has been observed that the $\group{z}[2]{}$ chiral critical bare mass in lattice units on $\Nt=6$ lattices is $\sim7.5$ times smaller than that found using $\Nt=4$.
Supposing a similar behavior also for the chiral tricritical mass $\muImRW$, then $\mTric[\text{light}]=0.043$ found on $\Nt=4$ lattices would imply $\mTric[\text{light}]=0.0057$ using $\Nt=6$.

\section{Analysis and preliminary results}

The procedure used to extract the critical exponent $\nu$ as well as the value of the kurtosis at $\betaC$ in the infinite volume limit is the same used in~\cite{Cuteri:2015qkq} and we refer to it for a detailed explanation.
The basic idea is to use the Ferrenberg-Swendsen reweighting~\cite{Ferrenberg:1989ui} to add few points between simulated $\beta$ and to fit simultaneously $\kurtosis(\beta,\Ns)$ on several spatial volumes linearly around $\betaC$.
In order to safely trust the outcome of the fit, finite size effects should not be dominant, i.e. the kurtosis measured on different volumes should meet at the same critical $\betaC$. 
Whenever this does not happen, a new larger spatial volume is needed and the smallest one should not be included in the finite size scaling analysis.
On the other hand, having data on three different $\Ns$ meeting at the same temperature, does not ensure the absence of any finite size correction and a check of the stability of the fit result leaving out the smaller volume is encouraged.

\begin{figure}[t]
    \centering
    \includegraphics[width=0.87\textwidth]{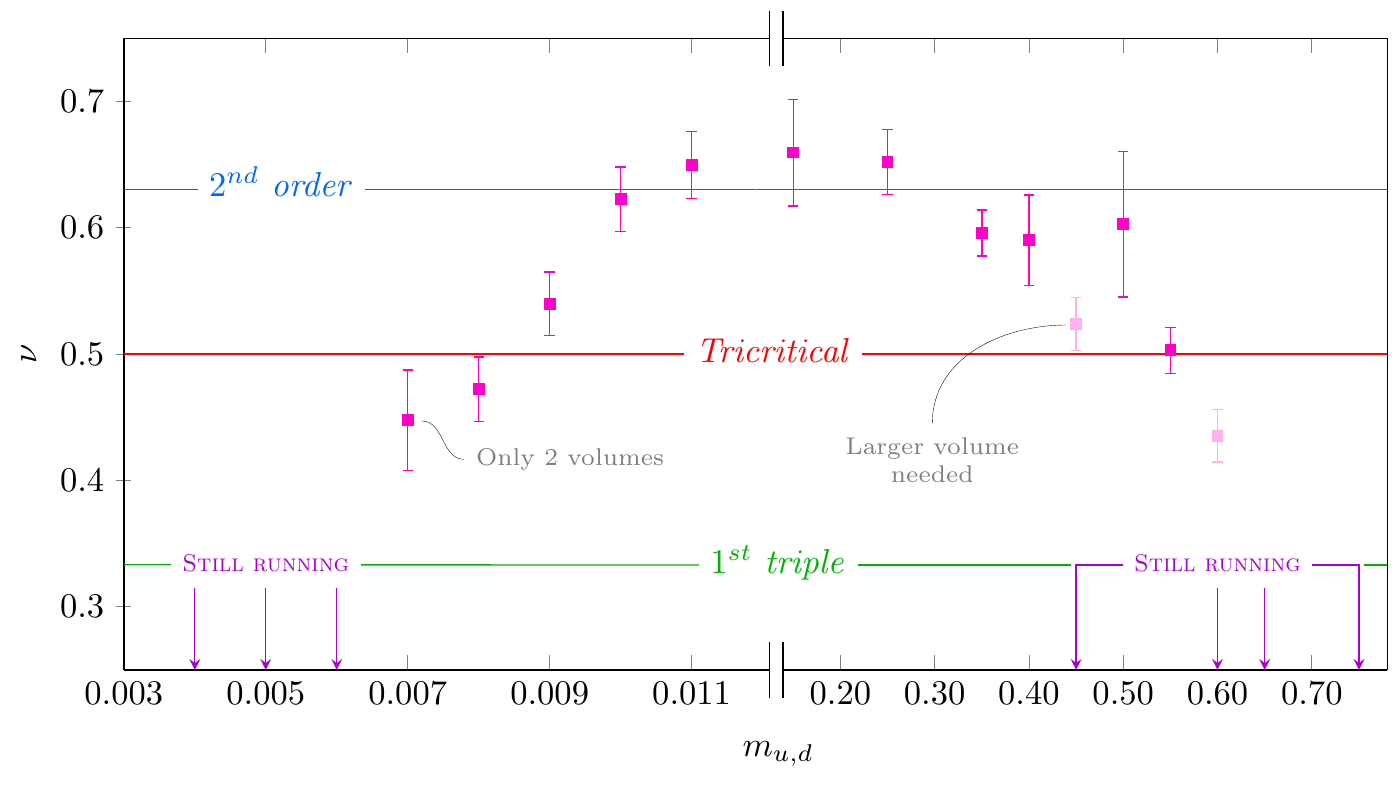}
    \caption{Critical exponent $\nu$ as function of the bare quark mass $\mass[]$.
             The horizontal colored lines are the critical values of $\nu$ for some universality classes.
             The mass axis has been broken and two different scales have been used in order to improve readability.
             Shaded points have to be considered preliminary.}
    \label{fig:nuVSmass}
\end{figure} 

In \cref{fig:nuVSmass} the extracted values of the critical exponent $\nu$ have been plotted for different bare quark masses.
As it has been already seen in previous studies~\cite{Bonati:2010gi,Cuteri:2015qkq}, leaving the \IIorder*{} region requires larger spatial volumes to be considered.
In particular, the smallest spatial lattice extent for the masses that can be considered finished has been $\Ns=12$ for $\mud\in[0.008,\,0.35]$, $\Ns=18$ for $\mud\in\{0.007,0.4\}$ and $\Ns=24$ for  $\mud=\{0.5,0.55\}$.
Despite some simulations are still running and, thus, some points are missing and some others are only preliminary, the expected behavior of $\nu$ can be seen.
In fact, a second order region for intermediate mass values separates two \Iorder*{} triple regions.
A first conservative estimate of the position of the two tricritical masses can be done,
\begin{equation*}
    \mTric[\text{light}]  = 0.008^{+0.002}_{-0.003} \qquad\text{and}\qquad
    \mTric[\text{heavy}] = \num{0.55 \pm 0.10} \quad.
\end{equation*}
The lack of points in the chiral region led us to choose an asymmetric error for the light tricritical mass. 
In the deconfinement region, instead, $\mud=0.45$ has been considered as lower bound for the tricritical point, even though the fit analysis gives a value of $\nu$ quite similar to the tricritical one.
This choice is due to the fact that more information can be gathered looking at the collapse plot of the kurtosis of the imaginary part of the \PL{}~--~i.e. plotting $\kurtosis(\PolyIm)$ as a function of the scaling variable $x\equiv(\beta-\betaC)\,\Ns^{1/\nu}$ at fixed $\betaC$ and $\nu$.
Since the data on the two largest volumes have a better collapse for $\nu=0.63$ than for $\nu=0.5$, we are confident that at this mass the system undergoes a \IIorder*{} phase transition in the thermodynamic limit and adding a larger volume will clarify the situation. 

\section{Conclusions and perspectives}

A preliminary comparison between the position of the tricritical masses on $\Nt=4$ and $\Nt=6$ lattices can be done in terms of $\mud/T=\Nt\:\latSpacing\mud$. 
\begin{table}[h]
    \centering
    \sisetup{table-number-alignment = center}
    %\begin{tabular}{*{2}{c}S[table-format = 1.6(2)]S[table-format = 1.3(2)]}
    \begin{tabular}{c@{\quad}S[table-format = 1.8(1)]@{\quad}S[table-format = 1.1(1)]}
        \toprule
        $\Nt$ & {$\mTric[\text{light}]/T$} & {$\mTric[\text{heavy}]/T$} \\
        \midrule
        4 & \num{0.172 \pm 0.020}       & \num{2.9 \pm 0.3} \\
        6 & {$0.048^{+0.012}_{-0.018}$} & \num{3.3 \pm 0.6} \\
        \bottomrule
    \end{tabular}
\end{table}
In the chiral region, the tricritical point moves towards smaller masses as already observed in previous studies with different discretizations, but the shift seems to be milder than at $\muRe=0$ with three degenerate flavors of staggered fermions.
$\mTric[\text{heavy}]/T$, instead, seems not to change and compatible values on $\Nt=4$ and $\Nt=6$ lattices are found.

Clearly, to draw any conclusion, some more work is needed.
In first place the missing points in \cref{fig:nuVSmass} will help to confirm the position of the tricritical masses.
Furthermore, the measurement of the pion mass and of the lattice spacing for each of the simulated quark bare masses is ongoing and this will allow a comparison with results obtained with different fermion discretizations, too.

Finally, the outcome of this study suggests that further investigations on even finer lattices in the chiral region are increasingly prohibitive.
The range of bare mass values to be simulated will probably further shift towards smaller values and it is known that this implies more costly simulations.
Moreover, increasing $\Nt$, larger spatial extents will be needed in order to perform the finite scaling analysis.
Consequently, new algorithmic improvements are required.
On the other hand, in the heavy mass region, it would be interesting to check whether the ratio $\mTric[\text{heavy}]/T$ stays constant even on finer lattices.
Even though this would still imply to simulate smaller bare quark masses and to use larger spatial volumes, it seems to be a feasible task.
However, it has to be checked if the lattice spacings used so far are fine enough to resolve the pion, i.e. whether $\latSpacing\mpi<1$.
Recent studies both at $\muRe=0$ and at $\muIm=\muImRW$ with Wilson fermions~\cite{Cuteri:2015qkq,Czaban:2016yae} have found that this is not the case and that in the deconfinement region larger values of $\Nt$ are needed.
Of course, different fermions discretizations are affected by different cut-off effects and the situation with staggered fermions could be different.
However, considering for instance past investigations on the equation of state for physical quark masses, it turned out that lattice temporal extents $\Nt\sim12$ where needed to perform a continuum extrapolation (see~\cite{Philipsen:2012nu} and references therein). 
Forthcoming results are going to clarify also this aspect.

\acknowledgments{We thank the staff of \lcsc{} for computer time and support.
                 This work is supported by the Helmholtz International Center for FAIR within the LOEWE program of the State of Hesse.}

\newpage

\bibliographystyle{JHEP}
\bibliography{Bibliography}

\end{document}